\documentstyle[floats,aps,prd,preprint,epsfig,eqsecnum,12pt]{revtex}
\tighten
\begin{document}
\preprint{
\parbox{45mm}{
\baselineskip=12pt
YCTP-P4-98\\
hep-th/9803225
\hspace*{1cm}
}}
\title{Towards Deriving Higgs Lagrangian from Gauge Theories}
\author{
 Noriaki Kitazawa\thanks{
  kitazawa@zen.physics.yale.edu}
  \thanks{On leave: Department of Physics Tokyo Metropolitan University
  Tokyo 192-03, Japan.}
 and
 Francesco Sannino\thanks{
  sannino@apocalypse.physics.yale.edu}
}
\address{Department of Phyiscs, Yale University,
                New Haven, CT 06520, USA}
\date{\today}
\maketitle
\begin{abstract}
A new method of deriving the Higgs Lagrangian
 from vector-like gauge theories
 is explored.
After performing a supersymmetric extension of gauge theories we 
identify  the auxiliary field associated with the ``meson'' superfield,
 in the low energy effective theory, as the composite Higgs field.
The auxiliary field, at tree level, has a ``negative squared mass''.
By computing the one-loop effective action
 in the low energy effective theory,
 we show that 
 a kinetic term for the auxiliary field emerges 
 when an explicit non-perturbative  
mechanism for supersymmetry breaking is introduced.
We find that, due to the  naive choice of the K\"ahler potential, 
 the Higgs potential remains unbounded from the below.  
A possible scenario for solving this problem is presented. 
It is also shown that
 once chiral symmetry is spontaneously broken
 via a non-zero vacuum expectation value of the Higgs field,
 the low energy composite fermion field acquires a mass and decouples, 
while in the supersymmetric limit it was kept massless 
by the 't Hooft anomaly matching conditions.
\end{abstract}
\draft

\section{Introduction}
\label{sec:intro}

Spontaneous chiral symmetry breaking
 induced by the fermion anti-fermion pair condensate
 is, in general, expected in asymptotically free vector-like gauge theories.
 The effective Higgs Lagrangian (linear $\sigma$-model) 
well describes chiral symmetry breaking at low energies, however 
 it is difficult to derive it from the original gauge theory.
It is known that  an effective Higgs Lagrangian can be derived from 
the Nambu-Jona-Lasinio model by using the so called auxiliary 
field method\cite{ER}.
An unsatisfactory feature is the fact that 
cannot be applied to gauge theories. 

There are already many attempts
 of deriving low energy effective theories associated with 
a given  gauge theory by  using supersymmetry  
\cite{MV,ASPY,DMS,EHS,MW,AM,ADKM,AMZ}.
The hope is to extend some of the  ``exact results'' deduced 
by Seiberg and Witten in \cite{S,SW} for 
supersymmetric vector-like gauge 
theories to non-supersymmetric gauge theories.

However, since in 
Ref.~\cite{MV,ASPY,DMS,EHS,MW,AM,ADKM,AMZ} 
the explicit supersymmetric breaking
 is treated perturbatively, supersymmetry cannot 
be completely decoupled.
Recently in Ref.~{\cite{SS,HSS}  
the issue of decoupling supersymmetry 
has been addressed, and a suitable decoupling 
procedure has been proposed for 
supersymmetric QCD like theories with $N_f<N_c$ and $N_c>2$.  
In the latter approach 
the decoupling is able to constrain only the holomorphic 
part of the QCD like potential which encodes the anomaly structure 
of the theory. A key point in the analysis of 
Ref.~{\cite{SS}} was to identify the 
auxiliary field associated with the supersymmetric effective 
low energy composite operators 
as the ``meson'' fields for the ordinary theory, once supersymmetry 
is broken.

In Refererence \cite{KS} we have outlined a new method for  
deriving the Higgs lagrangian.  In this paper we provide a more 
complete discussion of the method and we will also 
describe in some detail the new techniques 
for including non-perturbative supersymmetry breaking effects. 

Let us consider the $N=1$ supersymmetric extension
 for  vector-like gauge theories 
(assuming that the extension preserves asymptotic freedom).
We have the following quark chiral superfields for the underlying theory.
\begin{equation}
 Q^i_\alpha = \phi^i_{Q\alpha}
            + \sqrt{2} \theta \psi^i_{Q\alpha}
            + \theta^2 F^i_{Q\alpha},
 \qquad
 {\tilde Q}^{\alpha j} = \phi^{\alpha j}_{\tilde Q}
            + \sqrt{2} \theta \psi^{\alpha j}_{\tilde Q}
            + \theta^2 F^{\alpha j}_{\tilde Q},
\end{equation}
 where $i$ and $j$ are the flavor indices while $\alpha$ is the color index.
The low energy effective theory, including only massless fields, can be 
obtained by correctly reproducing the global  
symmetries and assuming holomorphy, as shown by Seiberg \cite{S}. 
The relevant massless chiral super field is 
the color-singlet ``meson'' field $M^{ij}$
 which couples to the quark bi-linear operator as follows
\begin{equation}
 M^{ij} \sim Q^i_\alpha {\tilde Q}^{\alpha j} \ ,
\end{equation}
where $\displaystyle{M=A + \sqrt{2} \theta \psi + \theta^2 F}$.
It is instructive to show how the different components in 
$M^{ij}$ couple to the operators defined in the underlying 
theory
\begin{eqnarray}
 A^{ij} &\sim& \phi_Q^i \phi_{\tilde Q}^j \ ,
\\
 \psi^{ij} &\sim& \psi_Q^i \phi_{\tilde Q}^j
                + \phi_Q^i \psi_{\tilde Q}^j \ ,
\\
 F^{ij} &\sim& \phi_Q^i F_{\tilde Q}^j + F_Q^i \phi_{\tilde Q}^j
             - \psi_Q^i \psi_{\tilde Q}^j \ ,
\end{eqnarray}
where the contraction of the color indices is understood.
If we consider massless quarks,
 the auxiliary field $F^{ij}$ couples only  to the quark bi-linear operator,
 because of the underlying field equation 
$\displaystyle{F_Q = F_{\tilde Q} = 0}$.
Of course, in the supersymmetric limit, 
the auxiliary field at the effective lagrangian level  
must be integrated out
 by using the algebraic equation of motion.

In order to investigate the effects of  the squark decoupling in the
 low energy physics, we introduce a soft supersymmetric 
breaking term which correctly 
reproduces the squark mass in the underlying theory.  
By using the spurion method  we add the following term
 in the K\"ahler potential 
\begin{equation}
 {\cal L}_{\rm soft}
 = \int d^4 \theta \, X \left(
                         Q^{\dag}Q + {\tilde Q}^{\dag} {\tilde Q}
                        \right) \ ,
\end{equation} 
of the fundamental theory. 
The spurion vector superfield $X$
 is a singlet under both gauge and chiral symmetry.
There are two possibilities for $X$:
 $\displaystyle{X = \theta^2 m + {\bar \theta}^2 m^{\dag}}$
 or $\displaystyle{X = - \theta^2 {\bar \theta}^2 m^2}$.
If we take the former,
 the equation of motion for the quark auxiliary field is modified with 
the respect to the supersymmetric limit, i.e. 
$\displaystyle{F_Q = - m A_Q}$ and 
$\displaystyle{F_{\tilde Q} = - m A_{\tilde Q}}$.
Therefore,
 the auxiliary field in the effective theory now 
 couples also to the squark bi-linear operator.
However when  $m$ is  large enough  compared 
with the dynamical generated 
scale $\Lambda$, we expect  the quark bi-linear operator
 to dominate over the squark one.
If we take the latter $X$ choice the equation of motion for the 
quark auxiliary field remains unchanged 
and the auxiliary field in the effective theory 
couples only to the quark bi-linear operator. 
Here we choose 
$\displaystyle{X=m\theta + m^{\dagger} \bar{\theta}}$, with 
$m$ real, as spurion. 

In this paper we assume the following naive K\"ahler potential
 for the effective ``meson'' superfield 
\begin{equation}
 K = {1 \over {(\alpha \Lambda)^2}} M^{\dag}_{ij} M^{ji} \ ,
\label{naive-K}
\end{equation}
where $\alpha$ is a numerical constant. 
This will automatically induce a  ``negative squared mass'' 
for the auxiliary field. 
At the component level we have 
\begin{equation}
 {\cal L}_{eff}
  = \int d^4 \theta \, K
  = {1 \over {(\alpha \Lambda)^2}}F^{\dag}_{ij} F^{ji} + \cdots \ ,
\end{equation}
 which means that
 the vacuum at $F=0$ is unstable
 and a non-zero vacuum expectation value
 (if  $F$ is regarded as a propagating field once supersymmetry 
is broken) is expected.
This fact supports our idea that the auxiliary field can be 
identified with the composite Higgs field associated with 
the quark bi-linear operator.  

It is known that the effective 
K\"ahler potential for $N=1$ supersymmetric theories 
cannot be fixed, so that a more general form 
is expected. 
However we believe that the K\"ahler potential 
 in Eq.~(\ref{naive-K})
 can be considered as the first term in a 
general low energy expansion \cite{K}.

The effect of supersymmetry breaking
 is incorporated in the effective theory by using the spurion method.
 The explicit supersymmetry breaking term in the K\"ahler potential
is of the form
\begin{equation}
 K_{soft} = {\beta \over {(\alpha\Lambda)^2}} 
X M^{\dag}_{ij} M^{ji} \ ,
\label{soft}
\end{equation}
 at the lowest order in a  $m/\Lambda$ expansion and 
$\beta$ is a numerical constant.

Once defined the K\"ahler potential in 
Eqs.~(\ref{naive-K}), (\ref{soft}) and added 
the appropriate exact superpotential \cite{S}, we 
then compute the one-loop effective action in 
the low energy effective theory.
It results in new terms to add to the K\"ahler potential. 
The loop calculation can also be interpreted as a way 
of providing more information about the K\"ahler potential 
in $N=1$ theories. 
If we had an exact K\"ahler potential 
(as for example in $N=2$ supersymmetry), we do not  
expect loop calculations to generate new contributions.

If we treat supersymmetry breaking effect perturbatively,
 the one-loop supergraph is infrared divergent. 
At the component level it is easy to recognize that the origin 
of such a divergence is the loop associated with 
the low energy effective scalar field.
   
 However,
 if we include the supersymmetry breaking effect non-perturbatively,
 the infrared divergence is automatically regulated 
 since the low energy scalar field becomes massive via $K_{soft}$.
The soft breaking mass, actually, works as an infrared cutoff.
Although it is possible to include a mass term  in the 
superpropagator for the effective super-field
 we prefer, for simplicity, to use the mass 
as an explicit infrared momentum cutoff
 within the Euclidean momentum cutoff scheme.

The paper is organized as follows. 
In section \ref{sec:model} we consider the asymptotically 
free $N=1$ supersymmetric gauge theory with number 
of colors $N_c=2$ and flavors $N_f=3$. 
We use this system as a laboratory
 to illustrate our new method of deriving 
the Higgs Lagrangian.
In section \ref{sec:one-loop}
 we compute the one-loop effective action for 
the low energy effective theory. By first computing 
the two point effective action we show that the 
kinetic term for the auxiliary field 
is naturally generated thus enforcing our idea of 
identifying the auxiliary field with the Higgs field.

We then demonstrate, by computing the four-point effective 
action contribution that, once chiral symmetry is spontaneously broken
 via a non zero vacuum expectation value of the Higgs field,
 the low energy effective fermion field acquires a mass term. 
In the supersymmetric limit the composite fermion was kept massless 
by the 't Hooft anomaly matching conditions {\cite{'tH}}. 
It is also seen that the Higgs potential remains unstable at the 
one loop level.   However 
by generalizing the K\"ahler potential 
we illustrate  a possible scenario for generating 
the complete Higgs potential.
In section \ref{sec:conclusion} we briefly review the main results 
and  consider possible improvements as well as some extension 
of the present model.

\section{A Simple Model}
\label{sec:model}
As  starting point we will consider 
the supersymmetric asymptotically free theory with 
$SU(2)$ gauge group and three flavors in the 
fundamental representation.  
The global quantum symmetry group is 
$SU(6) \otimes U(1)_R$, where $SU(6)$ is the enlarged 
flavor group associated with the quark chiral superfield 
$Q_\alpha^i$ and $U(1)_R$ is the $R$-symmetry. 
Here $i = 1,2,\cdots,6$ labels the flavor index
 while $\alpha = 1,2$ the color one.
We use the notation of Ref.\cite{WB} throughout this paper.
According to the 't Hooft anomaly matching conditions 
the ``meson'' chiral superfield
\begin{equation}
 M^{ij} \sim \epsilon^{\alpha\beta} Q_\alpha^i Q_\beta^j \ ,
\end{equation}
 which belongs to the $15_A$ representation of $SU(6)$, 
is a massless field.
By saturating at tree level the anomalous as well as 
non-anomalous Ward-Takahashi identities and imposing 
holomorphy \cite{S}  one can get the effective low energy 
superpotential 
\begin{equation}
 W_{eff} = - {1 \over {\Lambda^3}} {\rm Pf} M,
\end{equation}
 where $\Lambda$ is the dynamical scale associated with 
the underlying supersymmetric gauge theory.
To the ``exact'' super-potential we have the need, in 
order to completely define the theory, to add the 
K\"ahler potential. This is the most undetermined 
part for $N=1$ supersymmetric gauge theories 
in contrast with the $N=2$ supersymmetric theories 
\cite{SW} where the full K\"ahler is claimed to be known.  
Following the arguments presented in the introduction 
a reasonable approximation to the full K\"ahler potential 
would be to consider the following terms 
\begin{equation}
 K_{eff}
 = {1 \over {(\alpha\Lambda)^2}} M^{\dag}_{ij} M^{ji}
 + {\beta \over {(\alpha\Lambda)^2}} X M^{\dag}_{ij} M^{ji},
\end{equation}
 where $\alpha$ and $\beta$ are dimensionless numerical constants.
The need for a more general type of K\"ahler potential 
will be advocated 
later in the paper. 
The second term is a soft supersymmetry breaking term
 introduced using the spurion method,
 and provides a mass term for the scalar component
 of the effective superfield $M$.
We assume the spurion field to be of the form 
$X=m(\theta^2+{\bar \theta}^2)$ with real $m$. 
We can get the canonically normalized field by 
performing the following field rescaling
$M / (\alpha\Lambda) \rightarrow M$. 
The Lagrangian becomes 
\begin{equation}
 {\cal L}
 = \int d^4 \theta
   \left(
     M^{\dag}_{ij} M^{ji}
   + \beta X M^{\dag}_{ij} M^{ji}
   \right)
 - \alpha^3 \int d^2 \theta\, {\rm Pf} M + {\rm h.c.}.
\label{tree}
\end{equation}
By performing the ordinary theta integration 
we deduce the tree level 
``potential'' for the effective auxiliary field 
in the limit $\beta=0$
\begin{equation}
 V(F,A) = - F^{\dag}_{ij} F^{ji}
        + {{\alpha^3} \over {16}}
          \epsilon_{ijklmn} F^{ij} A^{kl} A^{mn}
        + {\rm h.c.},
\end{equation}
 where $F^{ij}$ and $A^{ij}$
 are, respectively, the auxiliary and the scalar components
 of the chiral superfield $M^{ij}$.
We observe that
 the auxiliary field has negative squared mass.
Since in the supersymmetric limit the auxiliary 
field does not propagate, i.e. no kinetic term 
is present in the tree level lagrangian, 
 it is not considered as a physical 
field and the instability is removed 
by integrating it out via its equation of motion
 $\partial V(F,A) / \partial F = 0$.
In the present approach, where supersymmetry 
is explicitly broken, we will keep it 
\cite{SS,KS} and will show next 
that via non-perturbative breaking effects 
a non trivial kinetic term is generated 
for the auxiliary field, suggesting that 
it can become a physical field.

\section{Calculation of the One-loop Effective Action}
\label{sec:one-loop}

In this section we present the actual calculations 
for the one-loop effective action based on the Lagrangian 
defined in Eq.~(\ref{tree}).
We employ the supergraph technique\cite{supergraph}  with the 
spurion $X$ as external field. 

\subsection{The Two Point Function}
\label{susec:two}
The first term to compute is the  
one-loop two point function which 
corresponds to the quadratic term in the effective 
field associated with the K\"ahler potential.

The diagram we evaluated is shown in Fig.\ref{two-point}.
By using the standard super Feynman rules we obtain
\begin{eqnarray}
 \Gamma_2
 = - 18 \left( {{\alpha^3} \over {2^3 3!}} \right)^2
     \int && {{d^4p} \over {(2\pi)^4}} {{d^4k} \over {(2\pi)^4i}}
             d^4 \theta_1 d^4 \theta_2 \,
   M^{\dag}_{ij}(-p,\theta_2)
    \, \epsilon^{ijbadc} \, \epsilon_{klabcd} \, M^{kl}(p,\theta_1)
\nonumber\\ &&
   {1 \over {k^2(p+k)^2}}
   \delta^4(\theta_2-\theta_1)
   {{{\bar D}^2(k,\theta_1) D^2(k,\theta_1)} \over {16}}
   \delta^4(\theta_1-\theta_2),
\end{eqnarray}
 where the factor $18$ is a symmetric factor
 and the factor $\alpha^3 / 2^3 3!$
 is the coupling constant defined in the superpotential.
With the  help of the following identity
\begin{equation}
 \delta^4(\theta_2-\theta_1)
 {{{\bar D}^2(k,\theta_1) D^2(k,\theta_1)} \over {16}}
 \delta^4(\theta_1-\theta_2)
 = \delta^4(\theta_2-\theta_1),
\label{D-formula-1}
\end{equation}
 we can integrate over $\theta_2$
 and deduce
\begin{equation}
 \Gamma_2
 = 18 \cdot (2 \cdot 4!) \left( {{\alpha^3} \over {2^3 3!}} \right)^2
   \int {{d^4p} \over {(2\pi)^4}} d^4 \theta \,
   M^{\dag}_{ij}(-p,\theta) M^{ji}(p,\theta)
   \int {{d^4k} \over {(2\pi)^4i}} {1 \over {k^2(p+k)^2}},
\end{equation}
 where we used the fact that  
\begin{equation}
 \epsilon^{ijabcd} \, \epsilon_{klabcd}
 = 4! \, \left( \delta^i_k \delta^j_l - \delta^i_l \delta^j_k \right).
\end{equation}
The $k$ momentum integration can be expressed as 
\begin{eqnarray}
 \int {{d^4k} \over {(2\pi)^4i}} {1 \over {k^2(p+k)^2}}
 &=& \int_0^1 dx \int {{d^4k} \over {(2\pi)^4i}}
                      {1 \over {\left[ k^2 + x(1-x) p^2 \right]^2}}
\nonumber\\
 &=& \int_0^1 dx \int_m^\Lambda {{d^4k_E} \over {(2\pi)^4}}
                      {1 \over {\left[ k_E^2 + x(1-x) p_E^2 \right]^2}}
\nonumber\\
 &\simeq&
     \int_0^1 dx \int_m^\Lambda {{d^4k_E} \over {(2\pi)^4}}
     {1 \over {k_E^4}}
     \left[ 1 - 2x(1-x) {{p_E^2} \over {k_E^2}} \right] 
\nonumber\\
 &\simeq&
     \int_m^\Lambda {{d^4k_E} \over {(2\pi)^4}} {1 \over {k_E^4}}
     - {1 \over 3} p_E^2
       \int_m^\Lambda {{d^4k_E} \over {(2\pi)^4}} {1 \over {k_E^6}} \ .
\end{eqnarray}
To perform the integral we have analytically continued it over  
the Euclidean region and assumed the scale $\Lambda$  as 
our ultraviolet cutoff.  This procedure corresponds to 
the Euclidean momentum cutoff regularization scheme. 
We remark that is seems natural to us to consider $\Lambda$ as 
a physical ultraviolet cutoff, since the present theory lives 
naturally below the scale $\Lambda$.

 The appearance of the infrared divergence is 
due to the scalar loop, hence 
 it is reasonable to identify the 
supersymmetric breaking mass $m$ with a physical 
infrared cutoff.
After regularizing the integral we performed a momentum 
expansion in $p_E^2$ with  $p_E^2 \ll m < \Lambda$.
It resulted in the following expression for the 
two point function
\begin{eqnarray}
 \Gamma_2
 &=&
   {3 \over {2\pi^2}} \left( {\alpha \over 2} \right)^6
   \int d^4 x d^4 \theta
   \left\{
    \ln \left( {{\Lambda^2} \over {m^2}} \right)
    M^{\dag}_{ij}(x,\theta) M^{ji}(x,\theta)
    + {1 \over {3m^2}} M^{\dag}_{ij}(x,\theta) \Box M^{ji}(x,\theta) 
   \right\}
\label{two-point-result}
\\
 &=&
   {3 \over {2\pi^2}} \left( {\alpha \over 2} \right)^6
   \int d^4 x d^4 \theta
   \left\{
    \ln \left( {{\Lambda^2} \over {m^2}} \right)
    M^{\dag}_{ij}(x,\theta) M^{ji}(x,\theta)
    + {1 \over {48m^2}}
      {\bar D}^2 M^{\dag}_{ij}(x,\theta) D^2 M^{ji}(x,\theta) 
   \right\} \ ,
\nonumber
\end{eqnarray}
 where
\begin{equation}
 M^{ij}(x,\theta)
 = \int {{d^4p} \over {(2\pi)^4}} M^{ij}(p,\theta) e^{-ipx}\ .
\end{equation}
The first term in the brackets of Eq.~(\ref{two-point-result})
 is a correction to the tree level K\"ahler potential while
 the second term is the source
 for the kinetic term of the auxiliary field.
By performing the usual theta integration for the second term 
we get 
\begin{equation}
 {\cal L}_{kin}
 = - {1 \over {2\pi^2}} \left( {\alpha \over 2} \right)^6
     {1 \over {m^2}} \partial_m F^{\dag}_{ij} \partial^m F^{ji}\ .
\end{equation}
We have shown that
 the auxiliary field really propagates and can be associated 
with a physical field.
It is worth stressing that
 this kind of non-perturbative supersymmetry breaking treatment 
in evaluating the momentum expansion 
 is actually essential for obtaining the previous result.
By canonically normalizing the Higgs field
\begin{equation}
 H \equiv {1 \over {\sqrt{2} \pi m}}
          \left( {\alpha \over 2} \right)^3 F \ ,
\label{higgs-def}
\end{equation}
 we deduce the following Lagrangian
\begin{equation}
 {\cal L}_2
 = - \partial_m H^{\dag}_{ij} \partial^m H^{ji}
   + \mu^2 H^{\dag}_{ij} H^{ji} \ ,
\end{equation} 
 where
\begin{equation}
 \mu^2 = m^2 \left[
              2 \pi^2 \left( {\alpha \over 2} \right)^6
              + 3 \ln \left( {{\Lambda^2} \over {m^2}} \right)
             \right] \ .
\end{equation}
Note that the one-loop contribution to the mass term 
for the Higgs field 
has the same sign than the tree level term.
We observe that the mass instability suggests that the 
Higgs can have a non-vanishing vacuum expectation value 
which would break the chiral symmetry group
 $SU(6)$ to $Sp(6)$.
In order to complete the picture we expect 
the existence of stabilizing terms 
in the Higgs potential. 
In the next section we estimate the 
fourth order term for the Higgs potential
 by employing the method outlined in this section.

\subsection{The Four Point Function}
\label{subsec:four}

The one-loop diagram for the four point function
 (quartic term for the effective field in the K\"ahler potential)
 is shown in Fig.\ref{four-point}.
The explicit evaluation of the Feynman diagram provides 
\begin{eqnarray}
 \Gamma_4
 &&=
   162 \left( {{\alpha^3} \over {2^3 3!}} \right)^4
   \epsilon_{klabcd}\epsilon^{ijabc'd'}
   \epsilon_{opa'b'c'd'}\epsilon^{mna'b'cd}
   \int {{d^4p} \over {(2\pi)^4}} {{d^4q} \over {(2\pi)^4}}
        {{d^4r} \over {(2\pi)^4}} {{d^4k} \over {(2\pi)^4i}}
\nonumber\\
 &&\int d^4 \theta_1 d^4 \theta_2 d^4 \theta_3 d^4 \theta_4 
   {{M^{kl}(p,\theta_1) M^{\dag}_{ij}(-(p+q+r),\theta_2)
     M^{op}(r,\theta_3) M^{\dag}_{mn}(q,\theta_4)}
    \over
    {k^2(k+p)^2(k+p+q)^2(k+p+q+r)^2}}
\nonumber\\
 &&\left(
    {{{\bar D}^2(k,\theta_1) D^2(k,\theta_1)} \over {16}}
    \delta^4 (\theta_1 - \theta_2)
   \right)
   \left(
    {{D^2(-(k+p+q),\theta_4) {\bar D}^2(-(k+p+q),\theta_4)} \over {16}}
    \delta^4 (\theta_3 - \theta_4)
   \right)
\nonumber\\
 &&\delta^4 (\theta_2 - \theta_3)
   \delta^4 (\theta_4 - \theta_1),
\end{eqnarray}
 where the factor $162$ is a symmetric factor.
After integrating over the variables $\theta_2$ and $\theta_4$
 and integrating by parts the $D$ and ${\bar D}$ operators
 with momentum $-(k+p+q)$, we have
\begin{eqnarray}
 \Gamma_4
 &=&
   162 \left( {{\alpha^3} \over {2^3 3!}} \right)^4
   \epsilon_{klabcd}\epsilon^{ijabc'd'}
   \epsilon_{opa'b'c'd'}\epsilon^{mna'b'cd}
   \int {{d^4p} \over {(2\pi)^4}} {{d^4q} \over {(2\pi)^4}}
        {{d^4r} \over {(2\pi)^4}} {{d^4k} \over {(2\pi)^4i}}
\nonumber\\
 &&\int d^4 \theta_1 d^4 \theta_3
   {{M^{\dag}_{ij}(-(p+q+r),\theta_3) M^{op}(r,\theta_3)}
    \over
   {k^2(k+p)^2(k+p+q)^2(k+p+q+r)^2}} \,
   \delta^4 (\theta_3 - \theta_1)
\nonumber\\
 &&\qquad
   \Bigg\{
    \left(
     {{{\bar D}^2 D^2} \over {16}}
     M^{\dag}_{mn}(q,\theta_1) M^{kl}(p,\theta_1) \right)
    \left(
     {{{\bar D}^2 D^2} \over {16}} \delta^4 (\theta_1 - \theta_3)
    \right)
\nonumber\\
 &&\qquad\qquad
  - 4
    \left(
     {{{\bar D}^{\dot\alpha} D^{\alpha}} \over 4}
     M^{\dag}_{mn}(q,\theta_1) M^{kl}(p,\theta_1)
    \right)
    \left(
     {{{\bar D}_{\dot\alpha} D_{\alpha}} \over 4}
     {{{\bar D}^2 D^2} \over {16}} \delta^4 (\theta_1 - \theta_3)
    \right)
\nonumber\\
 &&\qquad\qquad\qquad
  + M^{\dag}_{mn}(q,\theta_1) M^{kl}(p,\theta_1)
    \left(
     {{{\bar D}^2 D^2} \over {16}} {{{\bar D}^2 D^2} \over {16}}
     \delta^4 (\theta_1 - \theta_3)
    \right)
   \Bigg\} \ ,
\end{eqnarray}
 where for simplicity we omit the arguments of $D$ and ${\bar D}$.
By using the formulae
\begin{equation}
 \delta^4 (\theta_3 - \theta_1)
  {{{\bar D}_{\dot\alpha}(k,\theta_1) D_{\alpha}(k,\theta_1)} \over 4}
  {{{\bar D}^2(k,\theta_1) D^2(k,\theta_1)} \over {16}}
  \delta^4 (\theta_1 - \theta_3) 
  = {1 \over 2} \sigma^m_{\alpha{\dot\alpha}} k_m
    \delta^4 (\theta_3 - \theta_1)\ ,
\label{D-formula-2}
\end{equation}
\begin{equation}
 D^2(k,\theta_1) {{{\bar D}^2(k,\theta_1) D^2(k,\theta_1)} \over {16}}
 = - k^2 D^2(k,\theta_1),
\label{D-formula-3}
\end{equation}
together with Eq.~(\ref{D-formula-1}), we deduce 
\begin{eqnarray}
 \Gamma_4
 &=&
   162 \left( {{\alpha^3} \over {2^3 3!}} \right)^4
   \epsilon_{klabcd}\epsilon^{ijabc'd'}
   \epsilon_{opa'b'c'd'}\epsilon^{mna'b'cd}
   \int {{d^4p} \over {(2\pi)^4}} {{d^4q} \over {(2\pi)^4}}
        {{d^4r} \over {(2\pi)^4}} {{d^4k} \over {(2\pi)^4i}}
        d^4 \theta
\nonumber\\
 &&{1 \over {k^2(k+p)^2(k+p+q)^2(k+p+q+r)^2}} \,
   M^{\dag}_{ij}(-(p+q+r),\theta) M^{op}(r,\theta)
\nonumber\\
 &&\Bigg\{
     {{{\bar D}^2(p+q,\theta) D^2(p+q,\theta)} \over {16}}
   - {1 \over 2} \sigma^m_{\alpha{\dot\alpha}} k_m
      {\bar D}^{\dot\alpha}(p+q,\theta) D^{\alpha}(p+q,\theta)
   - k^2
   \Bigg\}
\nonumber\\
 &&M^{\dag}_{mn}(q,\theta_1) M^{kl}(p,\theta_1) \ .
\end{eqnarray}
Only the first term in the curly brackets 
gives the quartic term for the auxiliary field with no derivatives.
By keeping only the lowest order in momentum expansion 
we get
\begin{eqnarray}
 \Gamma_4'
 &=&
   162 \left( {{\alpha^3} \over {2^3 3!}} \right)^4
   \epsilon_{klabcd}\epsilon^{ijabc'd'}
   \epsilon_{opa'b'c'd'}\epsilon^{mna'b'cd}
   \int {{d^4p} \over {(2\pi)^4}} {{d^4q} \over {(2\pi)^4}}
        {{d^4r} \over {(2\pi)^4}}
        d^4 \theta
   \int {{d^4k} \over {(2\pi)^4i}} {1 \over {k^8}}
\nonumber\\
 &&M^{\dag}_{ij}(-(p+q+r),\theta) M^{op}(r,\theta)
   {{{\bar D}^2(p+q,\theta) D^2(p+q,\theta)} \over {16}}
   M^{\dag}_{mn}(q,\theta_1) M^{kl}(p,\theta_1) \ ,
\end{eqnarray}
The momentum integral can be estimated as in the
 previous section.
After contracting over the flavor indices and 
having performed the Fourier transformation, 
we have
\begin{eqnarray}
 \Gamma_4'
 &=&
   {1 \over {64\pi^2}} \left( {\alpha \over 2} \right)^{12}
   {1 \over {m^4}}
   \int d^4 x d^4 \theta
\nonumber\\
 &&\quad
   \left[
     M^{\dag}_{ij} M^{ji} {\bar D}^2 M^{\dag}_{kl} D^2 M^{lk}
   + M^{\dag}_{ij} M^{kl} {\bar D}^2 M^{\dag}_{lk} D^2 M^{ji}
   + 8 M^{\dag}_{ij} M^{jk} {\bar D}^2 M^{\dag}_{kl} D^2 M^{li}
   \right]\ .
\end{eqnarray}
This leads to the following contribution for the Higgs Lagrangian.
\begin{equation}
 {\cal L}_4
 = {{\pi^2} \over 8}
   \left[
    \left( H^{\dag}_{ij} H^{ji} \right)^2
    + 4 H^{\dag}_{ij} H^{jk} H^{\dag}_{kl} H^{li}
   \right]\ .
\label{H4-lowest}
\end{equation}
Unfortunately,
 the contribution is positive semi-definite
 and cannot stabilize the Higgs potential.
We attribute this negative result to the incompleteness 
of our naive K\"ahler potential. 
We also feel that the problem 
of deducing the full Higgs potential is 
as difficult as to completely solve the 
underlying gauge theory. 
While the effective superpotential is fixed, 
the K\"ahler potential is not for $N=1$ theories, 
hence, as we shall see, higher order K\"ahler terms 
are expected and can generate the correct Higgs potential.

Let us consider, for example, the effects of
the following K\"ahler term  
\begin{equation}
 K_{HE}
 = {\gamma \over {\Lambda^2}}
   \int d^4 x d^4 \theta 
   {\rm Tr} \left( M^{\dag} M M^{\dag} M \right)\ ,
\label{K-HE}
\end{equation}
 written for the canonically normalized effective field $M$,
 where $\gamma$ is a numerical constant.
The one-loop diagram for the four point function 
containing one vertex due to the above term leads to the 
following contribution for the Higgs potential
\begin{equation}
 {\cal L}_{HE}
 \sim \gamma {{m^2} \over {\Lambda^2}}
      \left( {\rm tr} \left( H^{\dag} H \right) \right)^2 \ .
\end{equation}
We observe  that this contribution might have the correct sign 
(depending on the sign of $\gamma$),
 while has different $m$-dependence
 with respect to the contribution in Eq.~(\ref{H4-lowest}).
On general grounds,
 if we were to consider all the high energy contributions to 
the K\"ahler potential,
 we expect the following form for the Higgs potential
\begin{equation}
 V_H = - m^2 f\left( {{m^2} \over {\Lambda^2}} \right)
         {\rm tr} \left( H^{\dag} H \right)
       + g\left( {{m^2} \over {\Lambda^2}} \right)
         \left( {\rm tr} \left( H^{\dag} H \right) \right)^2
       + h\left( {{m^2} \over {\Lambda^2}} \right)
         {\rm tr} \left( H^{\dag} H H^{\dag} H \right)\ ,
\end{equation}
where $f$, $g$ and $h$ are general functions of 
$m^2 /\Lambda^2$.
We can encode the obtained results in the previous functions as 
\begin{eqnarray}
 f\left( {{m^2} \over {\Lambda^2}} \right)
 &=& 2 \pi^2 \left( {\alpha \over 2} \right)^6
   + 3 \ln \left( {{\Lambda^2} \over {m^2}} \right) \ ,
\\
 g\left( {{m^2} \over {\Lambda^2}} \right)
 &=& - {{\pi^2} \over 8}
   + {\cal O} \left( {{m^2} \over {\Lambda^2}} \right) \ ,
\qquad
 h\left( {{m^2} \over {\Lambda^2}} \right)
 = - {{\pi^2} \over 2}
 + {\cal O} \left( {{m^2} \over {\Lambda^2}} \right)\ .
\end{eqnarray}
The vacuum expectation value for the Higgs field can be
written as
\begin{equation}
 v^2 = {{m^2 f(m^2/\Lambda^2)}
        \over
        {6 g(m^2/\Lambda^2) + h(m^2/\Lambda^2)}}\ ,
\end{equation}
where we assumed the breaking pattern, $SU(6)\rightarrow Sp(6)$, 
namely $\langle H \rangle = (v/\sqrt{2}) J$ with $J$ being the 
$Sp(6)$ invariant matrix normalized as $J^{\dag} J = 1$.
We conjecture that
 the complete non-perturbative calculation leads to a positive 
$v^2$ which persists in the limit $m \rightarrow \infty$.
The value of $v$ in this limit should be of the order of 
the new dynamical scale associated with the gauge 
theory \cite{SS} without squarks (i.e. 
quarks and a fermion in adjoint representation of the gauge group).

\subsection{The Four Point Function with One $X$ Insertion}
\label{subsec:four-x}

In this section we show that a non zero vacuum expectation 
value for the Higgs field triggers the appearance of a 
mass term for the effective fermion fields.
In the supersymmetric limit, since chiral symmetry 
is unbroken, the effective fermion fields are kept massless
 by the 't Hooft anomaly matching conditions.
However when chiral symmetry is spontaneously broken
 the anomalies are saturated via Nambu-Goldstone bosons
 and the fermions need not to be massless.   

Let us compute the four point function with one $X$ insertion 
whose Feynman diagram is provided in Fig.\ref{four-point-x}.
By evaluating the diagram we find
\begin{eqnarray}
 \Gamma_4^X
 &&=  324 \left( {{\alpha^3} \over {2^3 3!}} \right)^4
   \epsilon_{klabcd}\epsilon^{ijabc'd'}
   \epsilon_{opa'b'c'd'}\epsilon^{mna'b'cd}
   \int {{d^4p} \over {(2\pi)^4}} {{d^4q} \over {(2\pi)^4}}
        {{d^4r} \over {(2\pi)^4}} {{d^4k} \over {(2\pi)^4i}}
\nonumber\\
 &&\int d^4 \theta_1 d^4 \theta_2 d^4 \theta_3 d^4 \theta_4 d^4 \theta_5
   {{M^{kl}(p,\theta_1) \beta X(\theta_2) M^{\dag}_{ij}(-(p+q+r),\theta_3)
     M^{op}(r,\theta_4) M^{\dag}_{mn}(q,\theta_5)}
    \over
    {k^2 k^2 (k+p)^2 (k+p+q)^2 (k+p+q+r)^2}}
\nonumber\\
 &&\left(
    {{{\bar D}^2(k,\theta_1) D^2(k,\theta_1)} \over {16}}
    \delta^4 (\theta_1 - \theta_2)
   \right)
   \left(
    {{{\bar D}^2(k,\theta_2) D^2(k,\theta_2)} \over {16}}
    \delta^4 (\theta_2 - \theta_3)
   \right)
\nonumber\\
 &&\left(
    {{D^2(-(k+p+q),\theta_5) {\bar D}^2(-(k+p+q),\theta_5)} \over {16}}
    \delta^4 (\theta_4 - \theta_5)
   \right)
\nonumber\\
 &&\delta^4 (\theta_3 - \theta_4)
   \delta^4 (\theta_5 - \theta_1) \ ,
\end{eqnarray}
 where the factor $324$ is a symmetric factor.
After integrating over $\theta_2$, $\theta_4$ and $\theta_5$
 and using the relation
\begin{eqnarray}
  {{{\bar D}^2(k,\theta_1) D^2(k,\theta_1)} \over {16}}
&& X(\theta_1)
  {{{\bar D}^2(k,\theta_1) D^2(k,\theta_1)} \over {16}}
  \delta^4 (\theta_1 - \theta_3)
\nonumber\\
&& =- k^2
   \Bigg\{
    \left( {{{\bar D}^2} \over 4} X \right)
     {{D^2} \over 4}
   + {1 \over 2}
    \left( {\bar D}_{\dot{\alpha}} X \right)
     {\bar D}^{\dot{\alpha}} {{D^2} \over 4}
   + X {{{\bar D}^2 D^2} \over {16}}      
   \Bigg\}\delta^4 (\theta_1 - \theta_3) \ ,
\end{eqnarray}
 we obtain
 \begin{eqnarray}
 \Gamma_4^X
 &=&
 - 324 \left( {{\alpha^3} \over {2^3 3!}} \right)^4 
\beta 
\, \epsilon_{klabcd}\epsilon^{ijabc'd'}
   \epsilon_{opa'b'c'd'}\epsilon^{mna'b'cd}
   \int {{d^4p} \over {(2\pi)^4}} {{d^4q} \over {(2\pi)^4}}
        {{d^4r} \over {(2\pi)^4}} {{d^4k} \over {(2\pi)^4i}}
\nonumber\\
&& \int d^4 \theta_1 d^4 \theta_3
   {{M^{\dag}_{ij}(-(p+q+r),\theta_3) M^{op}(r,\theta_3)
     M^{\dag}_{mn}(q,\theta_1) M^{kl}(p,\theta_1)}
    \over
    {k^2 (k+p)^2 (k+p+q)^2 (k+p+q+r)^2}}
\nonumber\\
&& \quad
   \Bigg\{
    \left( {{{\bar D}^2(0,\theta_1)} \over 4} X \right)
     {{D^2(k,\theta_1)} \over 4}
   + {1 \over 2}
    \left( {\bar D}_{\dot{\alpha}}(0,\theta_1) X(\theta_1) \right)
     {\bar D}^{\dot{\alpha}}(k,\theta_1) {{D^2(k,\theta_1)} \over 4}
\nonumber\\
&& \qquad
   + X
  {{{\bar D}^2(k,\theta_1) D^2(k,\theta_1)} \over {16}} 
   \Bigg\} \delta^4 (\theta_1 - \theta_3)
\nonumber\\
&& \quad
   \left(
    {{D^2(-(k+p+q),\theta_1) {\bar D}^2(-(k+p+q),\theta_1)} \over {16}}
    \delta^4 (\theta_3 - \theta_1)
   \right) \ .
\end{eqnarray}
Integrating by parts the $D$ and ${\bar D}$ operators with momentum 
$-(k+p+q)$, 
while restricting our attention to the term which provides the fermion mass 
term, we have
\begin{eqnarray}
 \Gamma_4^X
 &\rightarrow& -324 \left( {{\alpha^3} \over {2^3 3!}} \right)^4
   \beta \, \epsilon_{klabcd}\epsilon^{ijabc'd'}
   \epsilon_{opa'b'c'd'}\epsilon^{mna'b'cd}
   \int {{d^4p} \over {(2\pi)^4}} {{d^4q} \over {(2\pi)^4}}
        {{d^4r} \over {(2\pi)^4}} {{d^4k} \over {(2\pi)^4i}}
\nonumber\\
&& \int d^4 \theta_1 d^4 \theta_3
   {{M^{\dag}_{ij}(-(p+q+r),\theta_3) M^{op}(r,\theta_3)
     }
    \over
    {k^2 (k+p)^2 (k+p+q)^2 (k+p+q+r)^2}}
\nonumber\\
&& \left(
    {{D^2(p+q,\theta_1)} \over 4}
    M^{\dag}_{mn}(q,\theta_1) M^{kl}(p,\theta_1)
   \right)
   \left(2\cdot {{{\bar D}^2(0,\theta_1)} \over 4} X(\theta_1) \right)
   \delta^4 (\theta_3 - \theta_1) \ ,
\end{eqnarray}
Now we can integrate over $\theta_3$
 and a straightforward calculation provides the following contribution
 to the effective Lagrangian
\begin{equation}
 {\cal L}^\psi_{mass}
 = - 2{\beta \over {\pi^2}} \left( {\alpha \over 2} \right)^{12}
     {1 \over {m^3}}
   \left\{
    F^{ij} F^{kl} \psi^{\dag}_{ij} \psi^{\dag}_{kl}
   + 4 F^{ij} F^{kl} \psi^{\dag}_{li} \psi^{\dag}_{jk}
   \right\},
\end{equation}
 where the contraction of the fermion indices is assumed.
In terms of the Higgs field defined in Eq.~(\ref{higgs-def}),
\begin{equation}
 {\cal L}^\psi_{mass}
 = - 4 \beta \left( {\alpha \over 2} \right)^6 {1 \over m}
   \left\{
    H^{ij} H^{kl} \psi^{\dag}_{ij} \psi^{\dag}_{kl}
   + 4 H^{ij} H^{kl} \psi^{\dag}_{li} \psi^{\dag}_{jk}
   \right\}.
\label{fermion-mass}
\end{equation}
We learn that the effective fermion field 
$\psi$ becomes massive once chiral symmetry is spontaneously 
broken via a non zero vacuum expectation value of the Higgs 
field $H$.

As for the Higgs potential the coefficient in 
Eq.~(\ref{fermion-mass}) becomes a general function 
of $m^2 / \Lambda^2$ once a 
more general K\"ahler is employed (see  Eq.~(\ref{K-HE})).
We then expect the effective fermion field to decouple as 
$m \rightarrow \infty$.

In this analysis we used as spurion the field 
$X = m ( \theta^2 + {\bar \theta}^2 )$.
In the introduction we have mentioned the possibility 
of realizing the spurion field as 
$X = - \theta^2 {\bar \theta}^2 m^2$.
However in order to generate a term like the one presented in 
Eq.~(\ref{fermion-mass}) with the new spurion field
 we need an odd number of operators $D^2$ and ${\bar D}^2$ 
acting on the effective fields $M$ and $M^{\dag}$ and $X$.
Within our approximation for the K\"ahler potential we 
cannot generate a fermionic type mass operator (at one-loop) if 
the latter choice of the the spurion field is made.  

\section{Conclusions}
\label{sec:conclusion}

We have explored a new method for deriving the Higgs Lagrangian
 from vector-like gauge theories.
The key idea is to identify the auxiliary field of the effective 
``meson'' superfield associated with the supersymmetrized 
version of the given gauge theory with the ordinary Higgs field. 
A similar identification was made in Ref.~\cite{SS,HSS} in 
order to define a suitable decoupling procedure 
from supersymmetric QCD to ordinary QCD like theories. 
In this paper we concentrated on providing a more
complete discussion of the novel method proposed 
in Ref.~\cite{KS} and we also described in some 
detail the new techniques for including non perturbative 
supersymmetric breaking effects.
  
We have then shown, by using as our laboratory the $SU(2)$ 
gauge theory with six matter doublets, that once 
supersymmetry is explicitly broken the auxiliary 
field, at the one loop level in the effective theory, 
acquires an ordinary kinetic term while the 
squared mass term already negative at the tree level 
remains negative at one loop. 
This supports the interpretation of the auxiliary 
field as the Higgs. 
We also remark that the present model illustrates  
how some of the non-holomorphic \cite{SS} terms for the 
non-supersymmetric low energy effective theory 
generates when supersymmetry 
decouples.
 
We have established that a better 
knowledge of the effective K\"ahler potential 
(associated with the full underlying dynamics) 
is essential to provide a bounded from 
the below Higgs potential. 

 We finally successfully generated 
a mass term for the effective fermion field 
which was kept massless in the supersymmetric limit
 by the 't Hooft anomaly matching conditions. 
An encouraging feature is that the non zero vacuum 
expectation value for the Higgs field, associated 
with spontaneous chiral symmetry breaking, triggers 
the appearance of the fermion mass.
In the resultant effective low energy theory 
the chiral anomalies are now saturated by the appearance of 
pseudoscalar Nambu-Goldstone bosons.

We remind the reader that determining the Higgs 
Lagrangian is as difficult as solving the complete 
dynamics in the underlying theory.  
In principle it is possible
 to extend the present  method for deriving the Higgs Lagrangian 
for arbitrary vector-like gauge theory.
However, there are some technical difficulties associated 
with the fact that the effective superpotential becomes 
highly involved for general $N_c$ and $N_f$. 
{}For example if we consider a $SU(N_c)$ gauge theory
 with $N_f = N_c + 1$ flavors 
we have to compute a ($N_f - 2$)-loop superdiagram
 to generate the kinetic term for the Higgs field.
Even more hard to handle are the $SU(N_c)$ gauge theories 
with $N_f \ne N_c + 1$. Indeed the associated supersymmetric low energy 
effective theories do not have a simple polynomial superpotential 
and is not straightforward to define a reliable loop calculation. 

It might be very interesting to generalize the present approach 
to the $N=2$ supersymmetric extension of gauge theories, 
since the supersymmetric low energy effective action 
(including the K\"ahler potential) in this case is completely 
known. However since the exact low energy effective theory is expressed,
for $N=2$ super theories, in terms of magnetic type variables 
the task would be to correctly identify the fermion 
condensate operator.

\acknowledgments

This work has been supported
 by the US DOE under contract DE-FG-02-92ER-40704.
The work by N.K. has also been supported in part
 by the Grant-in-Aid for Scientific Research 
 from the Ministry of Education, Science, and Culture of Japan
 on Priority Areas (Physics of CP violation)
 and \#09045036 under International Scientific Research Program,
 Inter-University Cooperative Research.

\begin{figure}
$$
\mbox{\epsfig{file=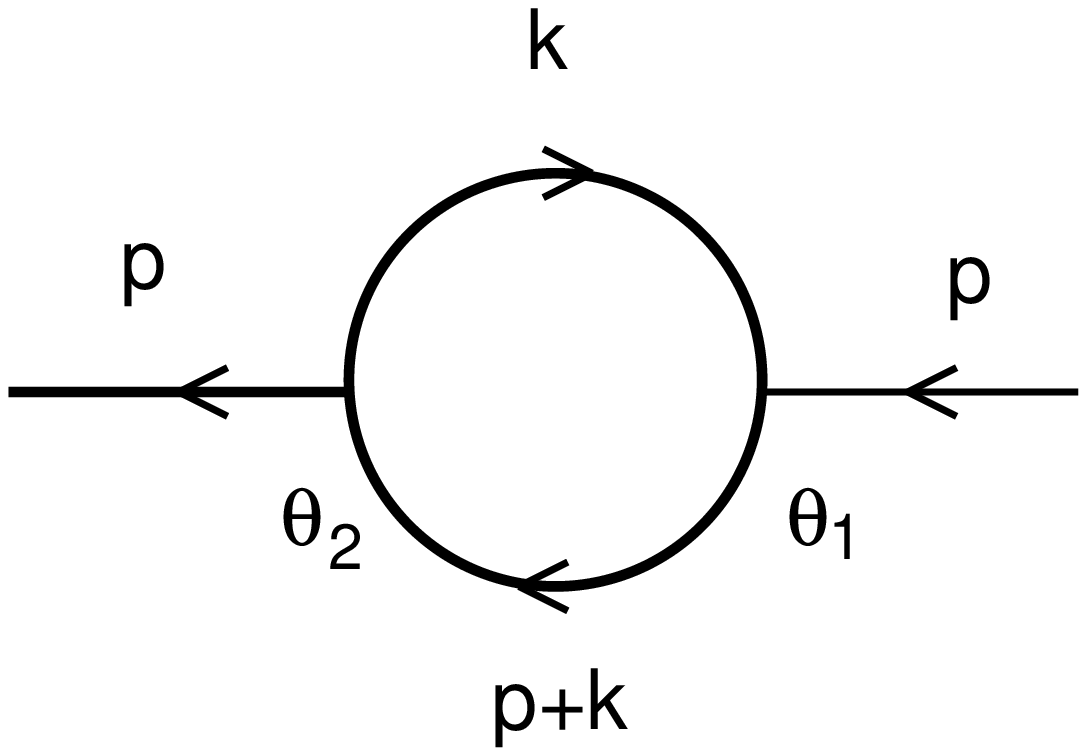,width=10cm}}
$$
\caption{The supergraph for the two point function.}
\label{two-point}
\end{figure}

\begin{figure}
$$
\mbox{\epsfig{file=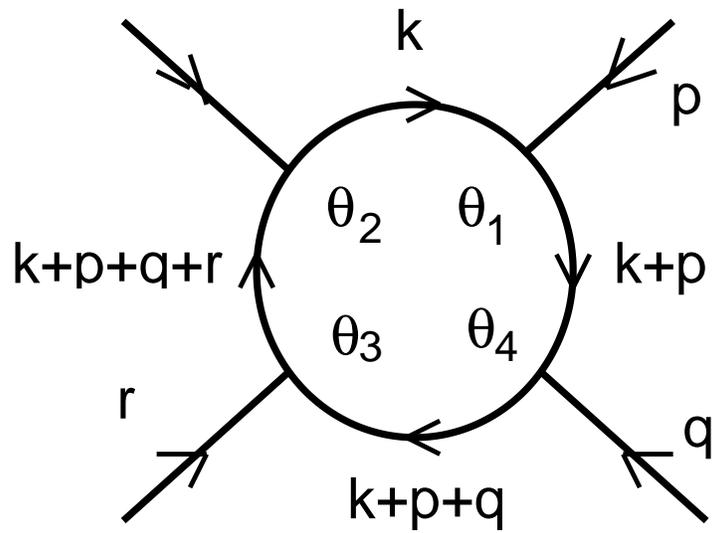,width=10cm}}
$$
\caption{The supergraph for the four point function.}
\label{four-point}
\end{figure}

\begin{figure}
$$
\mbox{\epsfig{file=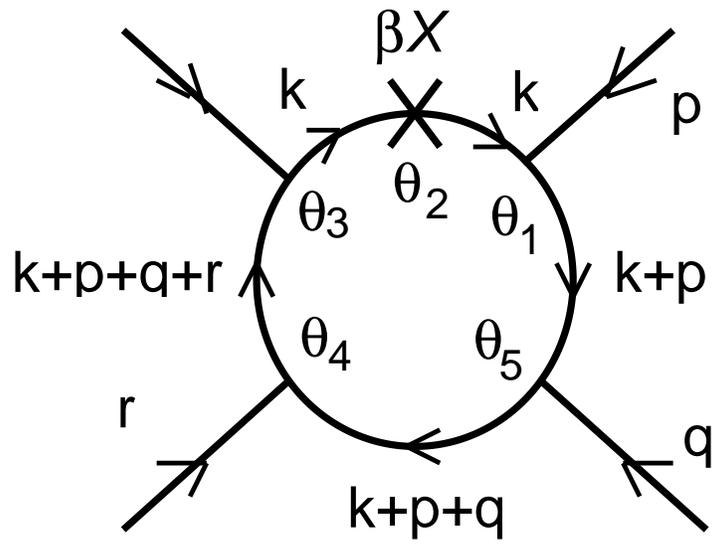,width=10cm}}
$$
\caption{The supergraph for the four point function
         with one $X$ insertion.}
\label{four-point-x}
\end{figure}

\end{document}